\documentclass{article}

\usepackage[lined,ruled,vlined]{algorithm2e}
\usepackage{amsmath,amsfonts,amssymb}
\usepackage{xspace}
\usepackage{graphicx}
\usepackage{xcolor}
\usepackage{amsthm}
\usepackage{geometry}

\newtheorem{thm}{Theorem}[section]

\newtheorem{definition}[thm]{Definition}

\newcommand{\R}{\ensuremath{\mathbb{R}}\xspace} 
\newcommand{\C}{\ensuremath{\mathbb{C}}\xspace}
\newcommand{\N}{\ensuremath{\mathbb{N}}}
\newcommand{\K}{\ensuremath{\mathbb{K}}} 
\newcommand{\M}{{\sf M}}
\newcounter{rmk} 
\newtheorem{remark}[rmk]{Remark}
\def\M{{\sf M}}

\usepackage{listings}
\lstnewenvironment{code}
{\lstset{language=C++,basicstyle=\scriptsize,  numbersep=6pt, frame=single, rulesep=.05cm,rulesepcolor=\color{black},showspaces=false,showstringspaces=false,moredelim=[is][\red]{|}{|}}}{}

\begin{document} 

\title{On Polynomial Multiplication in Chebyshev Basis}
\author{Pascal Giorgi
\thanks{P. Giorgi is with the Laboratoire d'Informatique, de Robotique et de Micro\'electronique de Montpellier (LIRMM), CNRS, Universit\'e Montpellier 2, 161 rue ADA, F-34095 Montpellier, France. E-mail: pascal.giorgi@lirmm.fr.}
}
\maketitle

\begin{abstract}
In a recent paper, Lima, Panario and Wang have provided a new method to multiply polynomials expressed in Chebyshev basis which reduces 
 the total number of multiplication for small degree polynomials. Although their method uses Karatsuba's multiplication, a quadratic number of operations is still needed. In this paper, we extend their result by providing a complete reduction to polynomial multiplication in monomial basis, which therefore offers many subquadratic methods.
Our reduction scheme does not rely on basis conversions and we demonstrate that it is efficient in practice. Finally, we show a linear time equivalence between the polynomial multiplication problem under monomial basis and under Chebyshev basis.
\end{abstract}
 
\section{Introduction }   
Polynomials are a fundamental tool in mathematics and especially in approximation theory
where mathematical functions are approximated using truncated series.
One can think of the truncated Taylor series to approximate a function as a polynomial expressed in monomial basis.
In general, many other series are preferred to the classical Taylor series in order to have better convergence properties. 
For instance, one would prefer to use the Chebyshev series  in order to have a rapid decreasing in the expansion coefficients which implies 
a better accuracy when using truncation \cite{Mason02, Boyd01}. One can also use other series such as Legendre or Hermite to achieve similar properties.
It is therefore important to have efficient algorithms to handle arithmetic on polynomials in such basis and especially for the multiplication problem \cite{CHEBFUN04,BJ2010}. 
 
Polynomial arithmetic has been intensively studied in the past decades, in particular following the work in 1962 of Karatsuba and Ofmann \cite{Kara62} who have shown that one can multiply  polynomials in a subquadratic number of operations. Let two polynomials of degree $d$ over a field $\K$ be given in monomial basis,  one can compute their product using Karatsuba's algorithm in $O(n^{\log_23})$ operations in $\K$. Since this seminal work, many other algorithms have been invented in order to asymptotically reduce the cost of the multiplication. In particular, one can go down to $O(n^{\log_{r+1}(2r+1)})$ operations in $\K$ with the generalized Toom-Cook method \cite{Toom63,Cook66} for any integer $r>0$. Finally, one can even achieve a quasi-linear time complexity using techniques based on the so-called FFT \cite{CooleyTukey65} (one can read \cite{VonzurGathen:1999:MCA} or \cite{MCA2010} for a good introduction and \cite{Cantor:1991:Kaltofen} for a study of fast polynomial multiplication over arbitrary algebras).
One of the main concern of this work is that all these algorithms have been designed for polynomials given in monomial basis, and they do not directly fit the other basis, such as the Chebyshev one. This is particularly true for the Karatsuba's method.

In this work, we extend the result of Lima, Panario and Wang \cite{LPW2010} who have partially succeeded in using Karatsuba's algorithm \cite{Kara62} within the multiplication of polynomials expressed in Chebyshev basis. Indeed, even if the method of \cite{LPW2010} uses Karatsuba's algorithm, its asymptotic complexity is still quadratic.
Our approach here is more general and it endeavors to completely reduce the multiplication in Chebyshev basis to the one in monomial basis. 
Of course, one can already achieve such a reduction by using back and forth conversions between the Chebyshev and the monomial basis using methods presented in \cite{BostanSalvySchost20008,BSS-2010}. However, this reduction scheme is not direct and it implies at least four calls to multiplication in monomial basis: three for the back and forth conversions and one for the multiplication of the polynomials. 
In this work, we present a new reduction scheme which does not rely on basis conversion and which uses only two calls to multiplication in monomial basis.
We also demonstrate that we can further reduces the number of operations by slightly modifying this reduction for the case of DFT-based multiplication algorithm. 
Considering practical efficiency, we will see that our reduction scheme will definitively compete with implementations of the most efficient algorithms available in the literature. 

{\it Organization of the paper}. Section \ref{sec:monomial} recalls some complexity results on polynomial multiplication in monomial basis and provides a detailed study on arithmetic operation count in the case of polynomials in $\R[x]$. In Section \ref{sec:chebmul} we give a short review on the available methods in the literature to multiply polynomials given in Chebyshev basis. Then, in Section \ref{sec:reduction} we propose our new method to perform such a multiplication by re-using multiplication in monomial basis. We analyze the complexity of this reduction and compare it to other existing methods. We perform some practical experimentations of such a reduction scheme in Section \ref{sec:expe}, and then compare its efficiency and give a small insight on its numerical reliability. Finally, we exhibit in Section \ref{sec:equivalence} the linear equivalence between the polynomial multiplication problem in Chebyshev basis and in monomial basis with only a constant factor of two.
 
\section{Classical Polynomial Multiplication }\label{sec:monomial}
It is well-known that polynomial multiplication of two polynomials in $\K[x]$ with degree  $d=n-1$ can be achieved with less than $O(n^2)$ operations in $\K$, for any field $\K$ (see \cite{VonzurGathen:1999:MCA, Cantor:1991:Kaltofen}), if polynomials are given in monomial basis.
Table \ref{tab:mularithcount} exhibits the arithmetic complexity of two well-known algorithms in the case of polynomials in $\R[x]$. One is due to Karatsuba and Ofman \cite{Kara62} and has an asymptotic complexity of $O(n^{\log_2 3})$ operations in $\K$; the other one is based on DFT computation using complex FFT and it has an asymptotic complexity of $O(n\log n)$ operations in $\K$, see \cite[algorithm 8.16]{VonzurGathen:1999:MCA} and \cite{Cantor:1991:Kaltofen,Schon71} for further details. One can see \cite{SplitRadixFFT2007,RealValuedFFT1987} for more details on complex FFT. We also give in  Table \ref{tab:mularithcount}  the exact number of operations in $\R$ for the schoolbook method. From now on, we will use $\log n$ notation to refer to $\log_2 n$.

\begin{table}[!ht] 
\begin{center}
\caption{Exact number of operations to multiply two polynomials over $\R[x]$ of degree $n-1$ in monomial basis with $n=2^k$}\label{tab:mularithcount}
\renewcommand{\arraystretch}{1.5}
\begin{tabular}{|l||c|c|}
\hline
\bf Algorithm & \bf nb. of multiplications & \bf nb. of additions \\
\hline 
Schoolbook    & $n^2$        & $(n-1)^2$         \\
\hline
Karatsuba & $n^{\log 3}$ &  $7n^{\log 3}-7n+2$ \\
\hline
DFT-based$^{(*)}$ & $3n \log 2n -4n+6$  & $9 n \log 2n -12n+12$  \\
\hline
\end{tabular}

\smallskip
{\it (*) using real-valued FFT of \cite{RealValuedFFT1987} with 3/3 strategy for complex multiplication}
\end{center}
\end{table}
To perform fast polynomial multiplication using DFT-based method on real inputs, one need to compute 3 DFT with $2n$ points, $n$ pointwise multiplications with complex numbers and $2n$ multiplications with the real constant $\frac{1}{2n}$. Note that we do not need to perform $2n$ pointwise multiplications since the DFT on real inputs has an hermitian symmetry property. Using Split-Radix FFT of \cite{RealValuedFFT1987} with  3/3 strategy for complex multiplication (3 real additions and 3 real multiplications), one can calculate the DFT with $n$ points of a real polynomial with $\frac{n}{2}\log n-\frac{3n}{2}+2$ real multiplications and $\frac{3n}{2} \log n-\frac{5n}{2}+4$ additions. Adding all the involved operations gives the arithmetic operation count given in Table \ref{tab:mularithcount}. Note that one can even decrease the number of operations by using the modified split-radix FFT of \cite{SplitRadixFFT2007}, yielding an overall asymptotic complexity of $\frac{34}{3}n\log 2n$ instead of $12n\log 2n$. 

In the following, we will use the function $\M(n)$ to denote the number of operations in $\R$ to multiply polynomials of degree less than $n$ when using the monomial basis. For instance, $\M(n)=O(n^{\log 3})$ with Karatsuba's algorithm. In order to simplify the notations, we assume throughout the rest of the paper that polynomials are of degree $d = n-1$ with $n=2^k$.

\section{Polynomial Multiplication in Chebyshev Basis}\label{sec:chebmul}

Chebyshev polynomials of the first kind on the interval $[-1,1]$ are defined by
\[
T_k(x)=\cos(k~{\rm arcos}(x)),\quad k\in\N^* \mbox{ and } x\in [-1,1]. 
\]  

According to this definition, one can remark that these polynomials are orthogonal polynomials. The 
following recurrence relation holds:

\[ 
\left\{
\begin{array}{l}
T_k(x)=2xT_{k-1}(x)-T_{k-2}(x)\\
T_0(x)=1\\
T_1(x)=x\\
\end{array}\right.
\]

It is obvious from this relation that the $i$-th Chebyshev polynomial $T_i(x)$ has degree $i$ in $x$.
Therefore, it is easy to show that $(T_i(x))_{i\geq 0}$ form a basis of the $\R$-vector space $\R[x]$.
Hence, every polynomial $f\in\R[x]$ can be expressed as a linear combination of $T_i(x)$. This representation is  called the Chebyshev expansion.
In the rest of this paper we will refer to this representation as the Chebyshev basis.\\

Multiplication in Chebyshev basis is not as easy as in the classical monomial basis.
Indeed, the main difficulty comes from the fact that the product of two basis elements spans over two other basis elements. The following relation illustrates this property:
\begin{equation}\label{eq:chebyprod}
T_i(x)~T_j(x)= \frac{T_{i+j}(x)+ T_{|i-j|}(x)}{2}, \quad \forall i,j\in\N.
\end{equation}

\subsection{Quadratic Algorithms}\label{sec:quadr}
According to (\ref{eq:chebyprod}), one can derive an algorithm to perform the product of two polynomials given in Chebyshev basis 
using a quadratic number of operations in $\R$. This method is often called the ``direct method''.
Let two polynomials $a,b\in\R[x]$ of degree $d= n-1$ expressed in Chebyshev basis :
\[
a(x)= \frac{a_0}{2}+\sum_{k=1}^{d}a_kT_k(x) \mbox{ and } b(x)= \frac{b_0}{2}+\sum_{k=1}^{d}b_kT_k(x).
\]
The $2d$ degree polynomial $c(x)=a(x)~b(x)\in\R[x]$ expressed in Chebyshev basis can be computed using the following formula  \cite{BasTas97}:

\[
c(x)= \frac{c_0}{2}+\sum_{k=1}^{2d}c_kT_k(x)
\]
such that
\begin{equation}\label{eq:directmethod}
 2c_k \hspace*{-.12cm} = \hspace*{-.12cm}
\begin{cases}
  a_0b_0+2\displaystyle\sum_{l=1}^{d}a_lb_l, & \hspace*{-.3cm}\text{for } k=0, \\[.5cm]
\displaystyle \sum_{l=0}^k \hspace*{-.04cm}a_{k-l}b_l+\hspace*{-.08cm}\sum_{l=1}^{d-k}\hspace*{-.04cm}(a_lb_{k+l}\hspace*{-.08cm}+a_{k+l}b_l), &\hspace*{-.3cm} \text{for } k=1,...,d\hspace*{-.08cm}-\hspace*{-.08cm}1, \\[.5cm]
\displaystyle \sum_{l=k-d}^{d} a_{k-l}b_l, & \hspace*{-.3cm}\text{for } k=d,...,2d.
\end{cases}
\end{equation}

The number of operations in $\R$ to compute all the coefficients of $c(x)$ using (\ref{eq:directmethod}) is exactly \cite{BasTas97,LPW2010}:
\begin{eqnarray}
&\bullet & n^2+2n-1 \mbox{ multiplications,}\nonumber\\
&\bullet & \displaystyle \frac{(n-1)(3n-2)}{2}  \mbox{ additions.}\nonumber
\end{eqnarray}

Lima {\it et al.} recently proposed in \cite{LPW2010} a novel approach to compute the coefficient of $c(x)$ which reduces the number of multiplications. 
The total number of operations in $\R$ is then: 
\begin{eqnarray}
&\bullet & \displaystyle \frac{n^2+5n-2}{2} \mbox{ multiplications,}   \nonumber \\
&\bullet & \displaystyle 3n^2+n^{\log 3}-6n+2 \mbox{ additions.} \nonumber
\end{eqnarray}

The approach in \cite{LPW2010} is to compute the terms $\sum a_{k-l}b_l$ using Karatsuba's algorithm \cite{Kara62} on polynomial $a(x)$ and $b(x)$ as if they were in monomial basis.

Of course, this does not give all the terms needed in~(\ref{eq:directmethod}). However, by reusing all partial results appearing along the recursive structure of Karatsuba's algorithm, the authors are able to compute all the terms $a_lb_{k+l}+a_{k+l}b_l$ with less multiplication than the direct method.
Even if the overall number of operations in $\R$ is higher than the direct method, the balance between multiplication and addition is different. 
The author claims this may have an influence on architectures where multiplier's delay is  much more expensive than adder's one.

\subsection{Quasi-linear Algorithms}\label{sec:quasilinear}

One approach to get quasi-linear time complexity is to use the discrete cosine transform (DCT-I).
The idea is to transform the input polynomials by using forward DCT-I, then perform a pointwise multiplication and finally transform the result back using backward DCT-I. An algorithm using such a technique has been proposed in \cite{BasTas97} and achieves a complexity of $O(n\log n)$ operations in $\R$.
As mentioned in \cite{LPW2010}, by using the cost of the fast DCT-I algorithm of \cite{ChanHo1990} one can deduce the exact number of operations in $\R$.
However, arithmetic operation count in \cite{LPW2010} is partially incorrect, the value should be corrected to: 
\begin{eqnarray}
&\bullet &  3n \log 2n-2n+3 \mbox{ multiplications},\nonumber\\
&\bullet & (9n+3)\log 2n-12n+12 \mbox{ additions}.\nonumber
\end{eqnarray} 

DCT-I algorithm of \cite{ChanHo1990} costs $\frac{n}{2}\log n-n+1$ multiplications and $\frac{3n}{2}\log n -2n+\log n+4$ additions when using $n$ sample points. To perform the complete polynomial multiplication, one needs to perform 3 DCT-I with $2n$ points, $2n$ pointwise multiplications and $2n$ multiplications by the constant $\frac{1}{2n}$. Adding all the operations count gives the arithmetic cost given above.

\section{Reduction To Monomial Basis Case} \label{sec:reduction}

\subsection{Using Basis Conversions}
One can achieve a reduction to the monomial basis case by converting the input polynomials given in Chebyshev basis to the monomial basis, then
perform the multiplication in the latter basis and finally convert the product back.
Hence, the complexity directly relies on the ability to perform the conversions between the Chebyshev and the monomial basis. 
In \cite{BSS-2010}, authors have proved that conversions between these two basis can be achieved in $O(\M(n))$ operations for polynomials of degree less than $n$. Assuming such reductions have a constant factor greater than or equal to one, which is the case to our knowledge, the complete multiplication of $n-$term polynomials given in Chebyshev basis would requires an amount of operation larger or equal to $4\M(n)$: at least $3\M(n)$ for back and forth conversions and $1\M(n)$ for the multiplication in the monomial basis. In the next section, we describe a new reduction scheme providing a complexity of exactly $2\M(n)+O(n)$ operations.

\subsection{Our Direct Approach}

As seen in Section \ref{sec:quadr}, Lima {\it et al.} approach \cite{LPW2010} is interesting since it introduces the use of monomial basis algorithms (i.e. Karatsuba's one) into Chebyshev basis algorithm.
The main idea in \cite{LPW2010} is to remark that the terms $\sum a_{k-l}b_l$  in (\ref{eq:directmethod}) are convolutions of order $k$. Hence, they are directly calculated in the product of the two polynomials 
\begin{eqnarray}\label{def:monomial} 
\displaystyle\bar{a}(x)&=&a_0+a_1x+a_2x^2+... +a_{d}x^{d},\nonumber\\[.2cm]
\displaystyle\bar{b}(x)&=&b_0+b_1x+a_2x^2+... +b_{d}x^{d}.
\end{eqnarray}
This product gives the polynomials 
\[
\displaystyle\bar{f}(x)=\bar{a}(x)~\bar{b}(x)=f_0+f_1x+f_2x^2+... +f_{2d}x^{2d}.
\]
Each coefficient $f_k$ of the polynomial $\bar{f}(x)$ corresponds to the convolution of order $k$.
Of course, this polynomial product can be calculated by any of the existing monomial basis algorithms (e.g. those of Section \ref{sec:monomial}). 
Unfortunately, this gives only a partial reduction to monomial basis multiplication. We now extend this approach to get a complete reduction.

Using coefficients $\bar{f}(x)$ defined above one can simplify (\ref{eq:directmethod}) to 
\begin{equation}\label{eq:directmethod2}
 2c_k = 
\begin{cases}
  f_0+2\displaystyle\sum_{l=1}^da_lb_l, & \text{for } k=0, \\[.5cm]
\displaystyle f_k+\sum_{l=1}^{d-k}(a_lb_{k+l}+a_{k+l}b_l), & \text{for } k=1,...,d-1, \\[.5cm]
\displaystyle f_k, & \text{for } k=d,...,2d.
\end{cases}
\end{equation}

In order to achieve the complete multiplication, we need to compute the three following summation terms for $k=1\hdots d-1$ : 
\begin{equation}\label{eq:term}
 \sum_{l=1}^d a_lb_l \mbox{ , }  \sum_{l=1}^{d-k} a_lb_{k+l} \mbox{ and }\sum_{l=1}^{d-k} a_{k+l}b_l.
\end{equation}

Let us define the polynomial $\bar{r}(x)$ as the reverse polynomial of $\bar{a}(x)$:
\[\bar{r}(x)=\bar{a}(x^{-1})x^d=r_0+r_1x+r_2x^2+\hdots+r_dx^d.\]
This polynomial satisfies $r_i=a_{d-i}$ for $i=0\hdots d$. 
Let the polynomial $\bar{g}(x)$ be the  product of the polynomials $\bar{r}(x)$ and $\bar{b}(x)$. Thus, we have
\[
\bar{g}(x)=\bar{r}(x)~\bar{b}(x)=g_0+g_1x+g_2x^2+\hdots+g_{2d}x^{2d}.
\]
The coefficients of this polynomial satisfy the following relation for $k=0\hdots d$ :
\[
\displaystyle g_{d+k}=\sum_{l=0}^{d-k}r_{d-l}b_{k+l}
\mbox{ and }
\displaystyle g_{d-k}=\sum_{l=0}^{d-k}r_{d-k-l}b_{l}.
\]
According to the definition of $\bar{r}(x)$ we have:
\begin{equation}\label{eq:reduction}
\displaystyle g_{d+k}=\sum_{l=0}^{d-k}a_{l}b_{k+l}
\mbox{ and }
\displaystyle g_{d-k}=\sum_{l=0}^{d-k}a_{k+l}b_{l}.
\end{equation}

All the terms defined in (\ref{eq:term}) can be easily deduced from the coefficients $g_{d+k}$ and $g_{d-k}$ of the polynomial $\bar{g}(x)$.
This gives the following simplification for (\ref{eq:directmethod2})

\begin{equation}\label{eq:directmethod3}
 2c_k\hspace*{-.12cm} = \hspace*{-.12cm}
\begin{cases}
\hspace*{-.08cm}f_0+2(g_d-a_0b_0), & \hspace*{-.2cm}\text{for } k=0, \\[.5cm]
\hspace*{-.08cm}f_k+g_{d-k}+g_{d+k}-a_0b_k-a_kb_0, & \hspace*{-.2cm}\text{for } k=1,...,d\hspace*{-.08cm}-\hspace*{-.08cm}1, \\[.5cm]
\hspace*{-.08cm}f_k, & \hspace*{-.2cm}\text{for } k=d,...,2d.
\end{cases}
\end{equation}

Applying (\ref{eq:directmethod3}), one can derive an algorithm which satisfies an algorithmic reduction to polynomial multiplication in monomial basis.
This algorithm is identified  as {\tt PM-Chebyshev} below.

\begin{algorithm}[!ht]
\SetKwInOut{Input}{Input}\SetKwInOut{Output}{Output}
\DontPrintSemicolon
\caption{ {\tt PM-Chebyshev}}\label{alg:pmchebyshev}
\Input{$a(x),b(x) \in \R[x]$ of degree $d= n-1$ with $\displaystyle a(x)= \frac{a_0}{2}+\sum_{k=1}^{d}a_kT_k(x)$ and $\displaystyle b(x)=  \frac{b_0}{2}+\sum_{k=1}^{d}b_kT_k(x)$.}
\Output{$c(x)\in \R[x]$ of degree $2d$ with
$\displaystyle c(x)=a(x)~b(x)=  \frac{c_0}{2}+\sum_{k=1}^{2d}c_kT_k(x).$}
\Begin{   
  let $\bar{a}(x)$ and $\bar{b}(x)$ as in (\ref{def:monomial})\\[.2cm]
  $\bar{f}(x):= \bar{a}(x)~\bar{b}(x)$\\[.2cm]
  $\bar{g}(x):= \bar{a}(x^{-1})x^d~\bar{b}(x)$\\[.2cm]
  $c_0:=\displaystyle\frac{f_0}{2}+g_d-a_0b_0$\\[.2cm]
  \For{$k=1$ {\bf to} $d-1$}{$\displaystyle c_k:= \frac{1}{2}(f_k+g_{d-k}+g_{d+k}-a_0b_k-a_kb_0)$}
  \For{$k=d$ {\bf to} $2d$}{$\displaystyle c_k:= \frac{1}{2}f_k$}
return $c(x)$
} 
\end{algorithm}

\subsection{Complexity Analysis} 

Algorithm {\tt PM-Chebyshev} is exactly an algorithmic translation of (\ref{eq:directmethod3}). Its correctness is thus immediate from (\ref{eq:directmethod2}) and (\ref{eq:reduction}). 

Its complexity is $O(\M(n))+O(n)$ operations in $\R$.
It is easy to see that coefficients $f_k$ and $g_k$ are computed by two products of polynomials of degree $d=n-1$ given in monomial basis.
This exactly needs $2\M(n)$ operations in $\R$. Note that defining polynomials $\bar{a}(x), \bar{b}(x)$ and $\bar{r}(x)$ does not need any operations in $\R$.
The complexity of the algorithm is therefore deduced from the number of operations in (\ref{eq:directmethod3}) and the fact that $d= n-1$.
The  exact number of operations in \R of Algorithm {\tt PM-Chebyshev} is $2\M(n)+8n-10$. The extra linear operations are divided into $4n-4$ multiplications and $4n-6$ additions.

Looking closely to (\ref{eq:term}) and (\ref{eq:reduction}), one can see that these equations only differ by the terms $a_0b_0$, $a_0b_k$ and $a_kb_0$.
This explains the negative terms in (\ref{eq:directmethod3}) which corrects these differences.
Assuming input polynomials have constant coefficients $a_0$ and $b_0$ equal to zero, then it is obvious that (\ref{eq:term}) and (\ref{eq:reduction}) will give the same values.
Considering this remark, it is possible to decrease the number of operations in Algorithm {\tt PM-Chebyshev} by modifying the value of the constant coefficient of $\bar{a}(x)$ and $\bar{b}(x)$ to be zero (i.e. $a_0=b_0=0$) just before the computation of $\bar{g}(x)$. Indeed, this removes all the occurrences of $a_0$ and $b_0$ in (\ref{eq:reduction}) which gives the following relation:
\begin{equation}
\displaystyle g_{d+k}=\sum_{l=1}^{d-k}a_{l}b_{k+l}
\mbox{ and }
\displaystyle g_{d-k}=\sum_{l=1}^{d-k}a_{k+l}b_{l},
\end{equation}
and therefore simplifies (\ref{eq:directmethod3}) to 
\begin{equation}
 2c_k\hspace*{-.12cm} = \hspace*{-.12cm}
\begin{cases}
  f_0+2g_d, & \hspace*{-.12cm}\text{for } k=0, \\[.5cm]
\displaystyle f_k+g_{d-k}+g_{d+k}, & \hspace*{-.12cm}\text{for } k=1,...,d-1, \\[.5cm]
\displaystyle f_k, & \hspace*{-.12cm}\text{for } k=d,...,2d.
\end{cases}
\end{equation}

Embedding this tricks into Algorithm {\tt PM-Chebyshev} leads to an exact complexity of $2\M(n)+4n-3$ operations in $\R$, where extra linear operations are divided into $2n-1$ multiplications and $2n-2$ additions.

\begin{table}[!ht]
\begin{center}
\caption{Arithmetic operation count in Algorithm {\tt PM-Chebyshev}}\label{tab:arithcount}
\renewcommand{\arraystretch}{1.5}
\begin{tabular}{|l||c|c|}
\hline
\bf \M(n) & \bf nb. of multiplication & \bf nb. of addition \\
\hline 
Schoolbook        & $2n^2+2n-1$        & $2n^2-2n$\\
\hline
Karatsuba         & $2n^{\log 3}+2n-1$  & $14n^{\log 3}-12n+2$\\
\hline
DFT-based$^{(*)}$ & $6n\log 2n-6n+11$      & $18n\log 2n-22n+22$\\
\hline
\end{tabular}

\medskip
{\it (*) using real-valued FFT of \cite{RealValuedFFT1987} with 3/3 strategy for complex arithmetic}
\end{center}
\end{table}

Table \ref{tab:arithcount} exhibits the exact number of arithmetic operation needed by Algorithm {\tt PM-Chebyshev} depending on the underlying algorithm  chosen to perform monomial basis multiplication. We separate multiplications from additions in order to offer a fair comparison to \cite{LPW2010} and we use  results in  Table \ref{tab:mularithcount} for $\M(n)$ costs.

\subsection{Special Case of DFT-based Multiplication}\label{sec:optimdft}

When using DFT-based multiplication, we can optimize the Algorithm {\tt PM-Chebyshev} in order to further reduce the number of operations. 
In particular, we can remark that Algorithm {\tt PM-Chebyshev} needs two multiplications in monomial basis using operands $\bar{a}(x), \bar{b}(x)$ and $\bar{a}(x^{-1})x^d,\bar{b}(x)$. Therefore, applying the generic scheme of Algorithm {\tt PM-Chebyshev}, we compute twice the DFT transform of $\bar{b}(x)$ on $2n$ points. 
The same remark applies to the DFT transform of  $\bar{a}(x)$ and $\bar{r}(x)=\bar{a}(x^{-1})x^d$ which can be deduced one from the other at a cost of a permutation plus $O(n)$ operations in $\R$.
Indeed, we have 
\hspace*{-.5cm}
\begin{eqnarray}
 &&\displaystyle {\tt DFT}_{2n}(\bar{a})=[~\bar{a}(w^k)~]_{k=0\hdots 2n-1},\nonumber\\
&&{\tt DFT}_{2n}(\bar{r})=[~\bar{a}(w^{-k})~\omega^{kd}~]_{k=0\hdots 2n-1}.\nonumber
\end{eqnarray}

Since $\omega=e^{\frac{-2i\pi}{2n}}$ by definition of the DFT, we have  $\omega^{2n}=1$ and therefore : 
\[\omega^k=\omega^{k-2n} \mbox{ and } \omega^{-k}=\omega^{2n-k} \mbox{ for } k\in\N.\]

This gives :
\[ 
{\tt DFT}_{2n}(\bar{r})=[~\bar{a}(w^{2n-k})~\omega^{dk}~]_{k=0\hdots 2n-1}.
\]
Considering the  DFT as an evaluation process, we have 
\[
\bar{r}(w^k)=(\omega_d)^k~\bar{a}(w^{2n-k}) \mbox{ for } k=0\hdots2n-1
\]
where $\omega_d=\omega^d=e^{\frac{-2i\pi d}{2n}}$. We can easily see that computing ${\tt DFT}_{2n}(\bar{r})$ 
is equivalent to reverse the values of ${\tt DFT}_{2n}(\bar{a})$ and multiply them by the adequate power of $\omega_d$.
This process needs exactly a permutation plus $4n-2$ multiplications in $\C$, which costs $O(n)$ operations while a complete FFT calculation needs $O(n \log n)$ operations.\\

Even if this method allows to decrease the number of operations, it needs extra multiplications with complex numbers which unfortunately makes the code numerically unstable (see the preliminary version of this work for a more detailed study \cite{Giorgi10}). As pointed out by one of the anonymous referees, this can be circumvent by slightly modifying the algorithm  {\tt PM-Chebyshev}.

Instead of computing $\bar{g}(x)=\bar{r}(x)~\bar{b}(x)$, it is more suitable to compute $\bar{h}(x)=\bar{s}(x)~\bar{b}(x)$ where $\bar{s}(x)=x~\bar{r}(x)$. Since $\bar{h}(x)$ has degree $2d+1$, which is equivalent to $2n-1$, it is still computable by applying a $2n$ points DFT-based multiplication. Following this, it suffices to shift coefficients of $\bar{h}(x)$ by one to get the coefficients of $\bar{g}(x)$ and then complete the algorithm. The trick here is that computing ${\tt DFT}_{2n}(\bar{s})$ is almost straightforward from ${\tt DFT}_{2n}(\bar{a})$ . Indeed, we have 
\[
{\tt DFT}_{2n}(\bar{s})=[~\bar{a}(w^{2n-k})~\omega^{nk}~]_{k=0\hdots 2n-1}. 
\]
Since $\omega$ is a $2n$-th primitive root of unity, it is obvious that $w^n=-1$ and thus we have
\[
{\tt DFT}_{2n}(\bar{s})=[~\bar{a}(w^{2n-k})~(-1)^{k}~]_{k=0\hdots 2n-1}.
\] 
This completely avoids the need to multiply the coefficients of ${\tt DFT}_{2n}(\bar{a})$ by power of $\omega_d$.
Using these considerations, one can modify Algorithm {\tt PM-Chebyshev} in order to save exactly the computation of 2 DFTs.
Hence, we obtain an arithmetic cost in this case of:
\begin{eqnarray}
&\bullet& 4n\log 2n+7 \mbox{ multiplications}, \nonumber\\
&\bullet& 12n\log 2n-12n+14 \mbox{ additions}.\nonumber
\end{eqnarray}

These values can be deduced by removing the cost of 2 DFT on $2n$ points in {\tt PM-Chebyshev} cost using DFT-based  multiplication (see Table \ref{tab:arithcount}). From now on, we will refer to this method as {\tt PM-Chebyshev} (DFT-based).\\

\begin{remark}
Instead of applying a permutation on the values of ${\tt DFT}_{2n}(\bar{a})$, one can use the hermitian symmetry property of real input DFT.
In other words, it is equivalent to say that $\bar{a}(\omega^{2n-k})$ is equal to the complex conjugate of $\bar{a}(\omega^{k})$.
This has no influences on the complexity analysis but considering real implementation it replaces memory swaps by modifications of the sign in the complex numbers structure. 
If data does not fit in cache, this might reduce the number of memory access and cache misses, and therefore provide better performances.\\
\end{remark}

\subsection{Comparisons With Previous Methods}\label{sec:compare} 

We now compare the theoretical complexity of our new method with existing algorithms presented in Section \ref{sec:chebmul}.

\begin{table}[!ht]
\begin{center}
\caption{Exact complexity for polynomial multiplication in Chebyshev basis, with degree $n-1$}\label{tab:totalarithcount}
\renewcommand{\arraystretch}{1.5}
\begin{tabular}{|l|l|}
\hline
\bf Algorithm & \bf nb. of operations in $\R$ \\
\hline 
\hline
Direct method                   & $2.5n^2-0.5n$ \\
\hline 
Lima {\it et al.} \cite{LPW2010} & $3.5n^2+n^{\log 3}-3.5n+1$ \\
\hline
DCT-based                       & $(12n+3)\log 2n-14n+15$ \\ 
\hline 
{\tt PM-Chebyshev} (Schoolbook)  & $4n^2-1$       \\
\hline
{\tt PM-Chebyshev} (Karatsuba)   & $16n^{\log 3}-10n+1$ \\
\hline
{\tt PM-Chebyshev} (DFT-based)   & $16n\log 2n-12n+21$ \\
\hline
\end{tabular}\label{tab:complexity}
\end{center} 
\end{table}   

In Table \ref{tab:complexity}, we report the exact number of operations in $\R$ for each methods.
One can conclude from this table that the asymptotically fastest multiplication is the one using DCT \cite{BasTas97}. However, according to the constants and the non-leading terms in each cost function, the DCT-based method is not always the most efficient, especially when polynomial degrees tend to be very small. Furthermore, we do not differentiate the cost of additions and multiplications which does not reflect the reality of computer architecture. 
For instance in the Intel\textregistered Core microarchitecture, which equipped the processor Intel Xeon 5330 launched in 2007, the latency for one double floating point addition is 3 cycles while it is 5 cycles for one multiplication \cite[table 2-15, page 2-35]{INTEL2010}. The same values apply to the Intel\textregistered Nehalem microarchitecture launched in 2010 \cite[table 2-19, page 2-50]{INTEL2010}.

\begin{figure}[!ht]
\caption{Theoretical speedup of polynomial multiplication in Chebyshev basis with different cost models.}\label{fig:costmodel}
\hspace*{-.5cm}\includegraphics*[width=.51\textwidth]{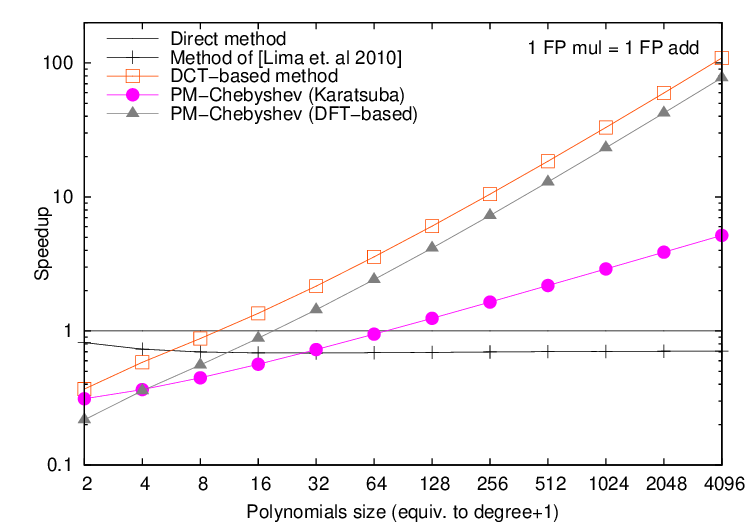}
\hspace*{-.5cm}\includegraphics*[width=.51\textwidth]{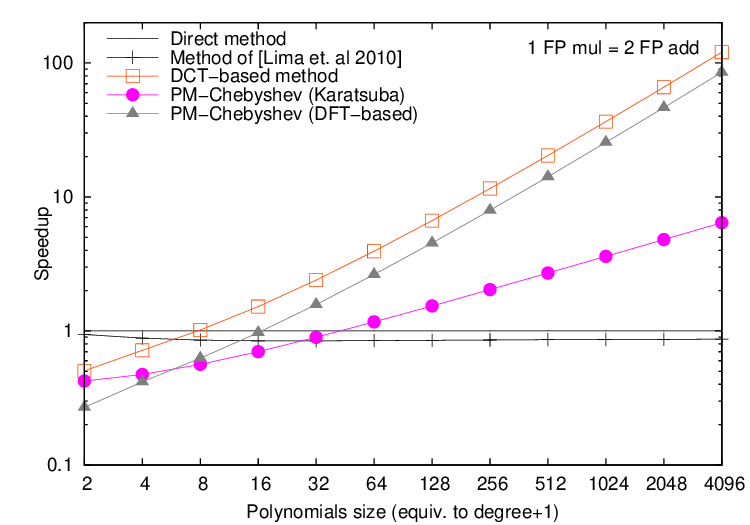}
\hspace*{-.5cm}\includegraphics*[width=.51\textwidth]{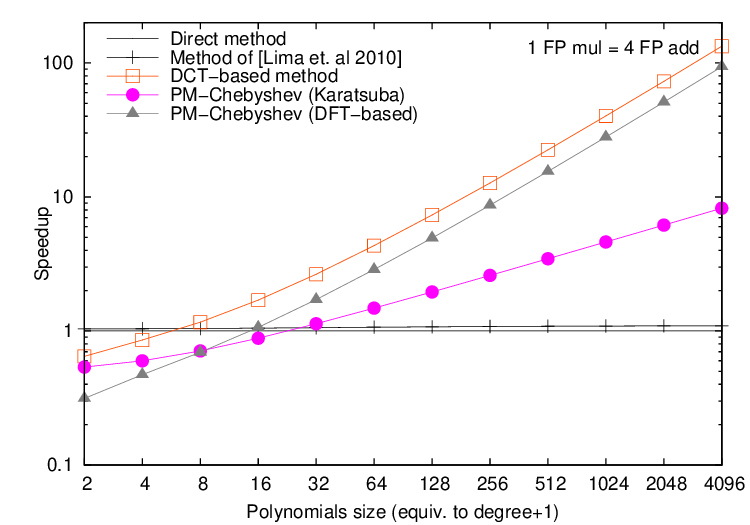}
\end{figure} 
   
In Figure \ref{fig:costmodel}, one can find the speedup of each methods compared to the Direct method. We provide different cost models to capture a little bit more the reality of nowadays computers where the delays of floating point addition and multiplication may differ by a factor of 4 at large.  Note that both axis use a logarithmic scale.

First, we can remark that changing cost model only affect the trade-off between methods for small polynomials (i.e. size less than 16).
As expected for large degrees, the DCT-based method is always the fastest and our Algorithm {\tt PM-Chebyshev} (DFT-based) is catching up with it since they mostly differ by a constant factor. However, when polynomial degrees tend to be small (less than 10) the Direct method is becoming the most efficient even if it has a quadratic complexity. 

As already mentioned in \cite{LPW2010}, the method of {\it Lima et al.} tends to become more efficient than the direct method for small polynomials when the cost model assumes that one floating point multiplication cost more than three floating point additions.
However, practical constraints such as recursivity, data read/write or cache access have an impact on performance, as we will see in Section \ref{sec:expe}, and need to be considered.

\section{Implementation and Experimentations}\label{sec:expe}  

In order to compare our theoretical conclusions with practical computations, we develop a software implementation of our Algorithm {\tt PM-Chebyshev}  and we report here its practical performances. As a matter of comparison, we provide implementations for previous known methods: namely the Direct method and the DCT-based method.
For the Direct method, a naive implementation with double loop has been done, while for the DCT-one we reuse existing software to achieve best possible performances.

\subsection{A Generic Code}
We design a C++ code to implement Algorithm {\tt PM-Chebyshev}  in  a generic fashion.
The idea is to take the polynomial multiplication in monomial basis as a template parameter in order to provide a generic function.
We decided to manipulate polynomials as vectors to benefit from the C++ Standard Template Library \cite{STL}, and thus benefit from genericity on coefficients, allowing the use of either double or single precision floating point numbers. Polynomial coefficients are ordered in the vector by increasing degree.
The code given in Figure \ref{fig:code} emphasizes the simplicity of our implementation:

\begin{figure}[!ht]\caption{Generic C++ code achieving the reduction to monomial basis multiplication.}\label{fig:code}
\begin{code}
template<class T, void mulM(vector<T>&,
                            const vector<T>&,
                            const vector<T>&)>
void mulC(       vector<T>& c,
           const vector<T>& a,
           const vector<T>& b){
  size_t da,db,dc,i;
  da=a.size(); db=b.size(); dc=c.size();
 
  vector<T> r(da),g(dc);
 
  for (i=0;i<da;i++) 
     r[i]=a[da-1-i];
 
  mulM(c,a,b); 
  mulM(g,r,b);
 
  for (i=0;i<dc;++i) 
     c[i]*=0.5;
 
  c[0]+=g[da-1]-a[0]*b[0];
 
  for (i=1;i<da-1;i++)
     c[i]+= 0.5*(g[da-1+i]+g[da-1-i]-a[0]*b[i] -a[i]*b[0]);	
}
\end{code}
\end{figure}
\smallskip
The function {\tt mulM} corresponds to the implementation of the multiplication in monomial basis while the function {\tt mulC}  corresponds to the one in Chebyshev basis. The vectors {\tt a} and {\tt b} represent the input polynomials and {\tt c} is the output product.
As expected, this code achieves a complete reduction to any implementation of polynomial multiplication in monomial basis, assuming the prototype of the function is compliant. In our benchmarks, we will use this code to reduce to a homemade code implementing the recursive Karatsuba's multiplication algorithm. Our Karatsuba's implementation is using a local memory strategy to store intermediate values along the recursive calls and allocations are done directly through {\tt new/delete} mechanism.
We also use a threshold to switch to naive quadratic product when polynomial degrees are small. In our benchmark, the threshold has been set to degree $63$ since it was the most efficient value.

\subsection{Optimized Code Using DCT and DFT}\label{sec:dctdft}
Many groups and projects have been already involved in designing efficient implementations of discrete transforms such as DCT and DFT. We can cite for instance the Spiral project \cite{SPIRAL} and  the FFTW library effort \cite{FFTW05}.
In order to benefit from the high efficiency of these works, we build our DCT/DFT based codes on top of the FFTW routines. 
For both DCT and DFT computations we use FFTW plans with FFTW\_MEASURE planning option, which offer optimized code using runtime measurement of several transforms. 

As explained in the documentation of the FFTW library, the DCT-I transform using a pre/post processed real DFT suffers from numerical instability. Therefore, the DCT-I implementation in FFTW is using either a recursive decomposition in smaller optimized DCT-I codelets or a real DFT of twice the size plus some scalings.
For the latter case, this means that the complexity of the DCT-I code is not reflecting the one of \cite{ChanHo1990} we used in our complexity analysis.
Taking this into account, one should replace the $2n$ points DCT-I transforms of Section \ref{sec:quasilinear} by $4n$ points DFT transforms plus $2n$ multiplications by the real constant $2$.
This increases the complexity of the DCT-based method to :
\begin{eqnarray}
&\bullet &  6n \log 4n-12n+6 \mbox{ multiplications},\nonumber\\
&\bullet & 18n\log 4n-30n+12 \mbox{ additions}. \nonumber
\end{eqnarray} 

\begin{figure}[!ht]  
\centering
\caption{Theoretical speedup of polynomial multiplication in Chebyshev basis.}
\hspace*{-.5cm}
\includegraphics*[width=0.7\textwidth]{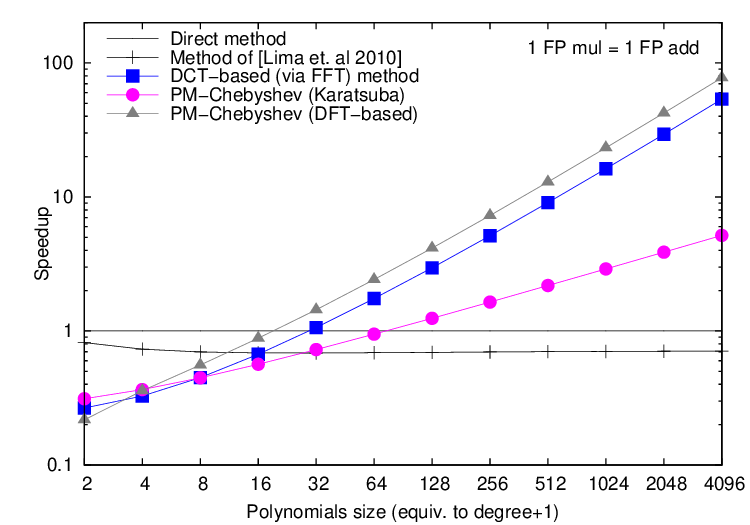} \label{fig:compare-arith-total} 
\end{figure} 
Considering this practical stability issue and its impact on the complexity of the DCT-based, we can see in Figure \ref{fig:compare-arith-total} the effect on the theoretical speedups.
In particular, our DFT-based reduction is always more efficient than the DCT-based, especially when polynomial degrees are large. This is of course
explained by the difference of the constant term in the complexity: $16 n\log 2n$ for our method and $24 n\log 2n$ for the DCT-based (via FFT).

\subsection{Code Validation}
As a matter of reliability, we check the validity of all our implementations.
First, we check their correctness by verifying the results of their implementations done in a symbolic way using Maple\footnote{www.maplesoft.com} software.

Since we want to perform numerical computations,  it is clear that the accuracy of the results may differ from one method to another.
It is therefore crucial to investigate their stability to give good statement on the accuracy.
It is not the intend of this work to give statements on the accuracy and this task would definitively require a dedicated work.
However, in order to give a small insight  we did some experiments to emphasize the relative error of every methods. 
Let us now give the definition of the relative error on polynomials as given in \cite{Higham2002}. \\

\begin{definition}
Let $a(x),b(x)$ be polynomials given in Chebyshev basis with double precision floating point numbers coefficients.
We define $\hat{c}(x)$ to be the approximation of the product $a(x)b(x)$ using double precision computation (53 bits of mantissa) and $c(x)$ to be the exact product computed over rational numbers. 
Using this notation,  the relative error $E(\hat{c}(x))$ is defined as 
\[
E(\hat{c}(x))=\frac{\lVert c(x)-\hat{c}(x) \rVert_2}{\lVert c(x) \rVert_2}
\]
where  $\lVert\hdots\rVert_2$ represents the Euclidean norm of polynomials, i.e. $\lVert a(x)\rVert_2= (\sum_{k=0}^d a_k^2)^\frac{1}{2}$ where the $a_k$ correspond to the coefficients of $a(x)$.\\
\end{definition} 

\begin{figure}[!ht]
\centering
\caption{Experimental measure of the relative error in double precision (Intel Xeon 2GHz). Entries lying in $[-50,50]$}
\hspace*{-.4cm}\includegraphics*[width=0.7\textwidth]{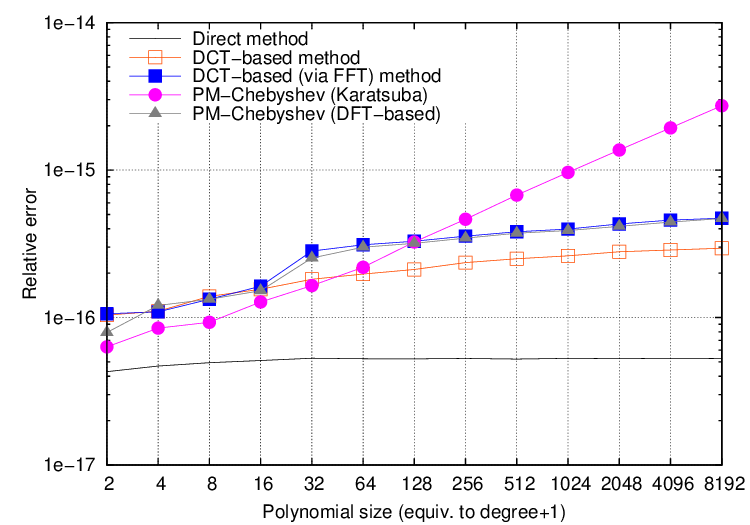}\label{fig:precision_intel}  
\end{figure}

Following this definition, we have computed the relative error on polynomial products using polynomial inputs having random floating point entries. 
While the numerical results are computed in double precision floating point numbers, the exact product is computed using the arbitrary precision rational numbers of the GMP\footnote{http://gmplib.org/} library.
The relative error is almost computed exactly since only the square root is using floating point approximations, the remaining parts being computed over the rationals.
We propose in Figure \ref{fig:precision_intel}  the measure of the relative error in our experiments.
The ordinates axis gives the average relative error of $50$ products with different random double precision floating point entries lying between $-50$ and $50$. 

\begin{figure}[!ht] 
\centering
\caption{Experimental measure of the relative error in double precision (Intel Xeon 2GHz). Entries lying in $[0,50]$}
\hspace*{-.4cm}\includegraphics*[width=0.7\textwidth]{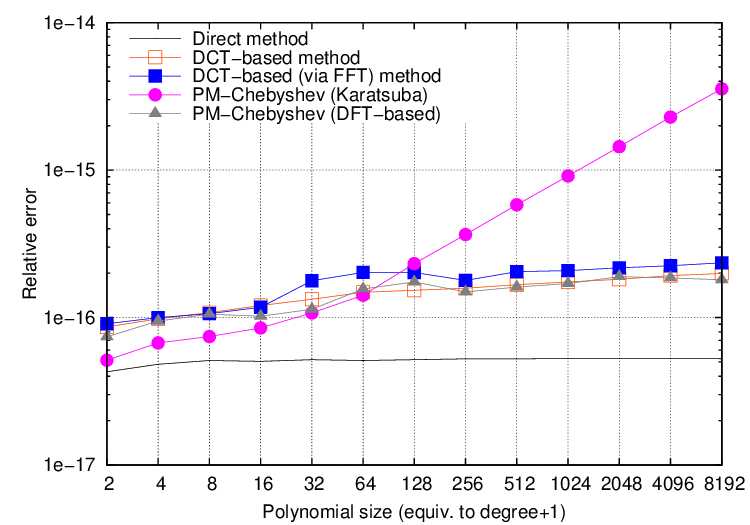}\label{fig:precision_intel2} 
\end{figure}

One can see in this figure that Algorithm {\tt PM-Chebyshev} (DFT-based) seems to offer the same numerical quality as its DCT-based counterparts.
This can be explained by the fact that both methods shared the same computational core (i.e. DFT transforms) and no other sensible numerical operations are performed.
Even if we change a little bit the settings of our experiment, as in Figure \ref{fig:precision_intel2} where we consider only positive floating point random entries (e.g. in $[0,50]$), the numerical stability of Algorithm {\tt PM-Chebyshev} (DFT-based) is still catching up with the one of DCT-based methods.

However, as soon as Karatsuba method is used in Algorithm {\tt PM-Chebyshev}, the stability is decreasing according to the growth of polynomial degree. This can be motivated by the nature of Karatsuba method which replaces one multiplication by few additions, and thus may introduce more numerical errors.

From these experiments we can conclude few thoughts.
Algorithm {\tt PM-Chebyshev} (DFT-based) seems to offer the same numerical behavior as the DCT based method, and thus offer a concrete alternative in practice as it will increase efficiency (see section \ref{sec:benchmark}). If one prefers to use Algorithm {\tt PM-Chebyshev}(Karatsuba), one
has to be careful with the given results since its numerical behavior sounds more unstable.
Finally, a theoretical study of the numerical stability of all these methods has to be done to give precise statements on their reliability.

\subsection{Benchmarks} \label{sec:benchmark}
 
We now compare the practical efficiency of the different methods.
We performed our benchmarks on an architecture which represents nowadays processors: an Intel Xeon processor 5130 running at 2GHz with $2\times$4MB of L2 cache. We use the gcc compiler version 4.4.5 with O3 optimization. Even if the platform is multi-core, we did not use any parallel computations and the FFTW library has been built sequential. 
For each method, we  measure the average running time of several polynomial multiplications. All the computations have been done with double precision floating point numbers and with the same data set. 

\begin{remark}
We only offer an average running time estimate of each algorithms since it is not realistic on nowadays processor to estimate precise running time of computation taking few milliseconds. 
\end{remark}
 
\begin{figure}[!ht]  
\centering
\caption{Practical performances of polynomial multiplication in Chebyshev basis against direct method - Intel Xeon 2GHz (global view). }
\hspace*{-.5cm}\includegraphics*[width=0.5\textwidth,angle=-90]{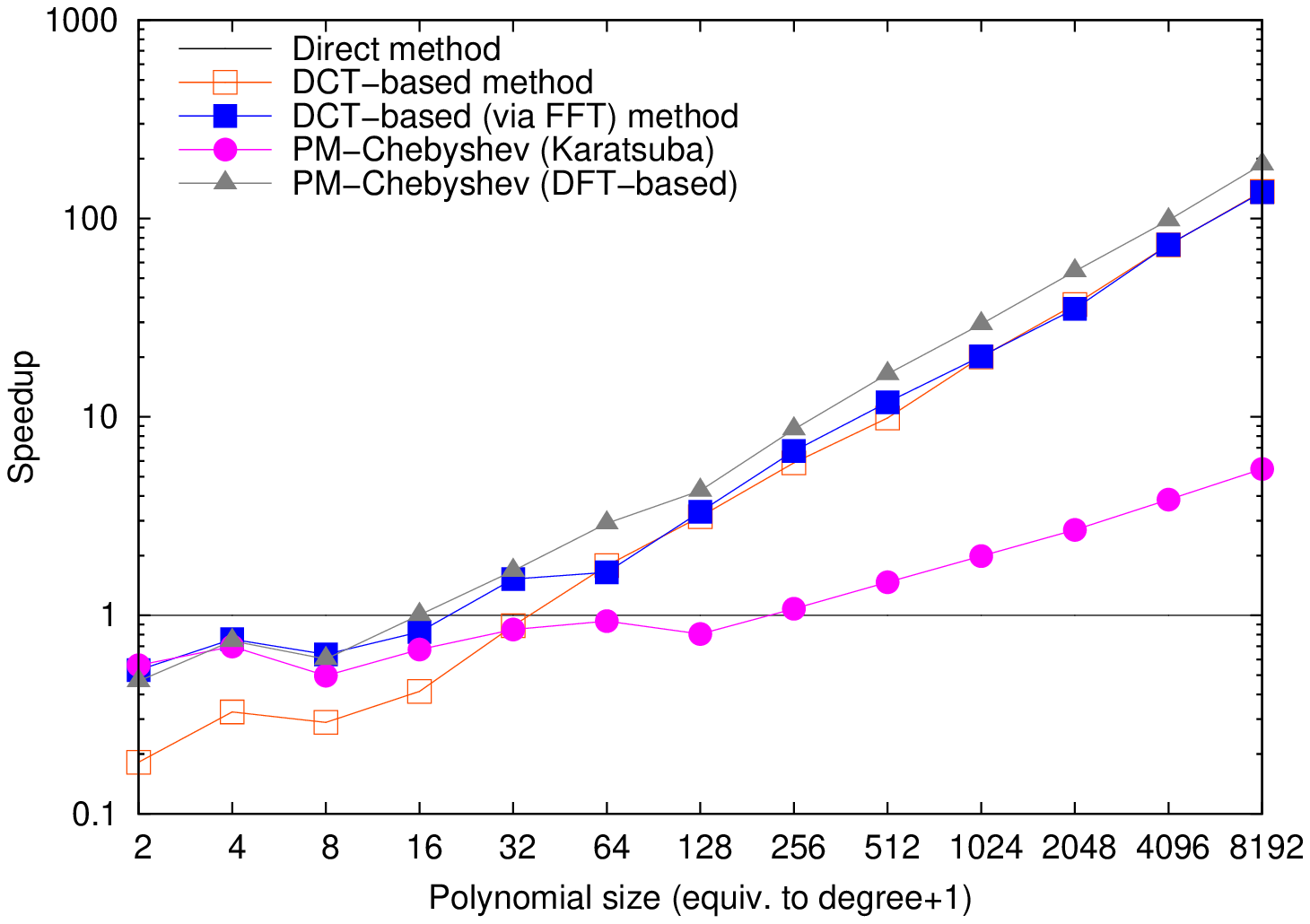}\label{fig:perf_intel}
\end{figure}

\begin{figure}[!ht]
\caption{Practical performances of polynomial multiplication in Chebyshev basis against direct method - Intel Xeon 2GHz (partial view).}\label{fig:partialview}
\hspace*{-.5cm}\includegraphics*[width=0.36\textwidth,angle=-90]{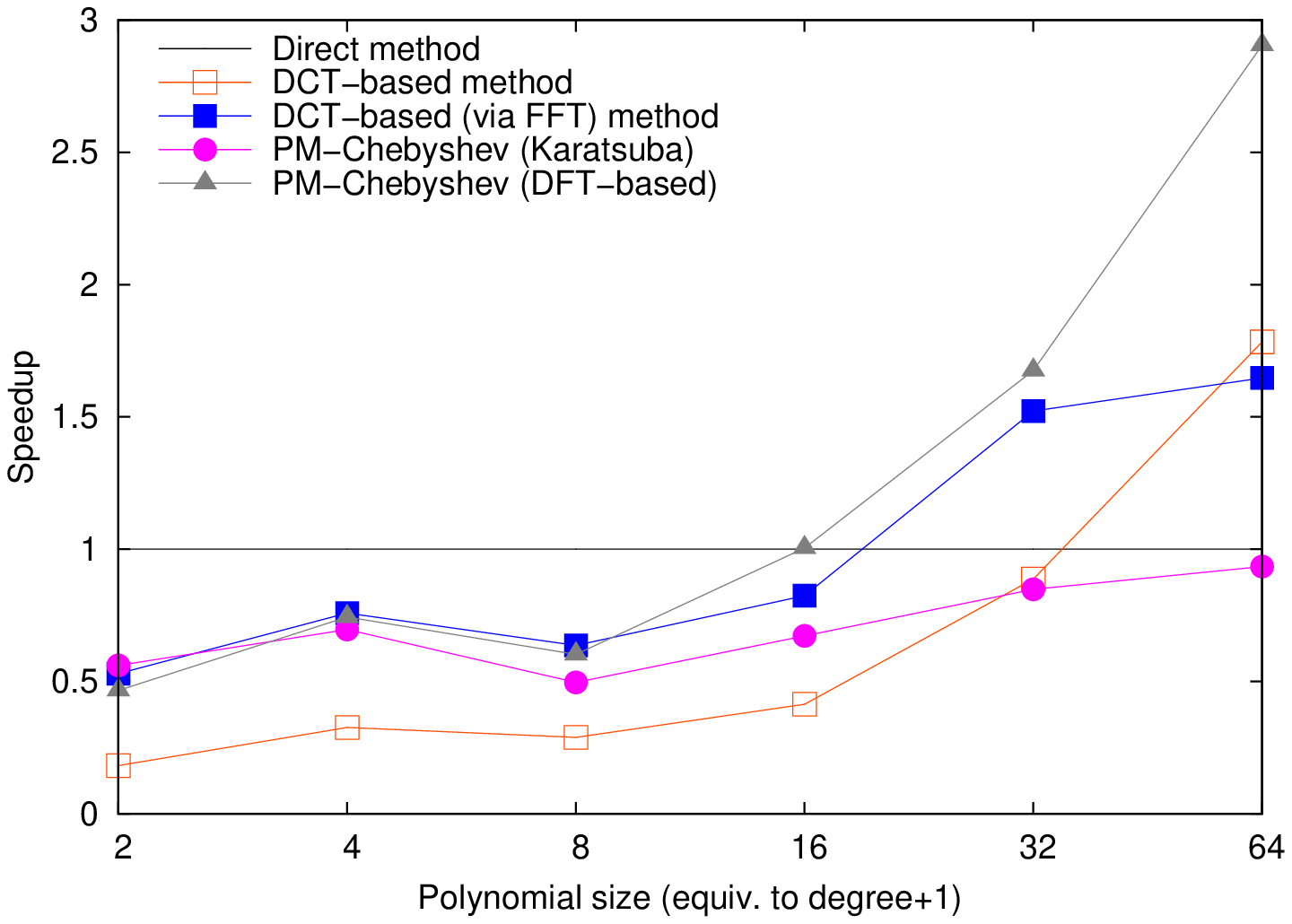}
\hspace*{-.5cm}\includegraphics*[width=0.36\textwidth,angle=-90]{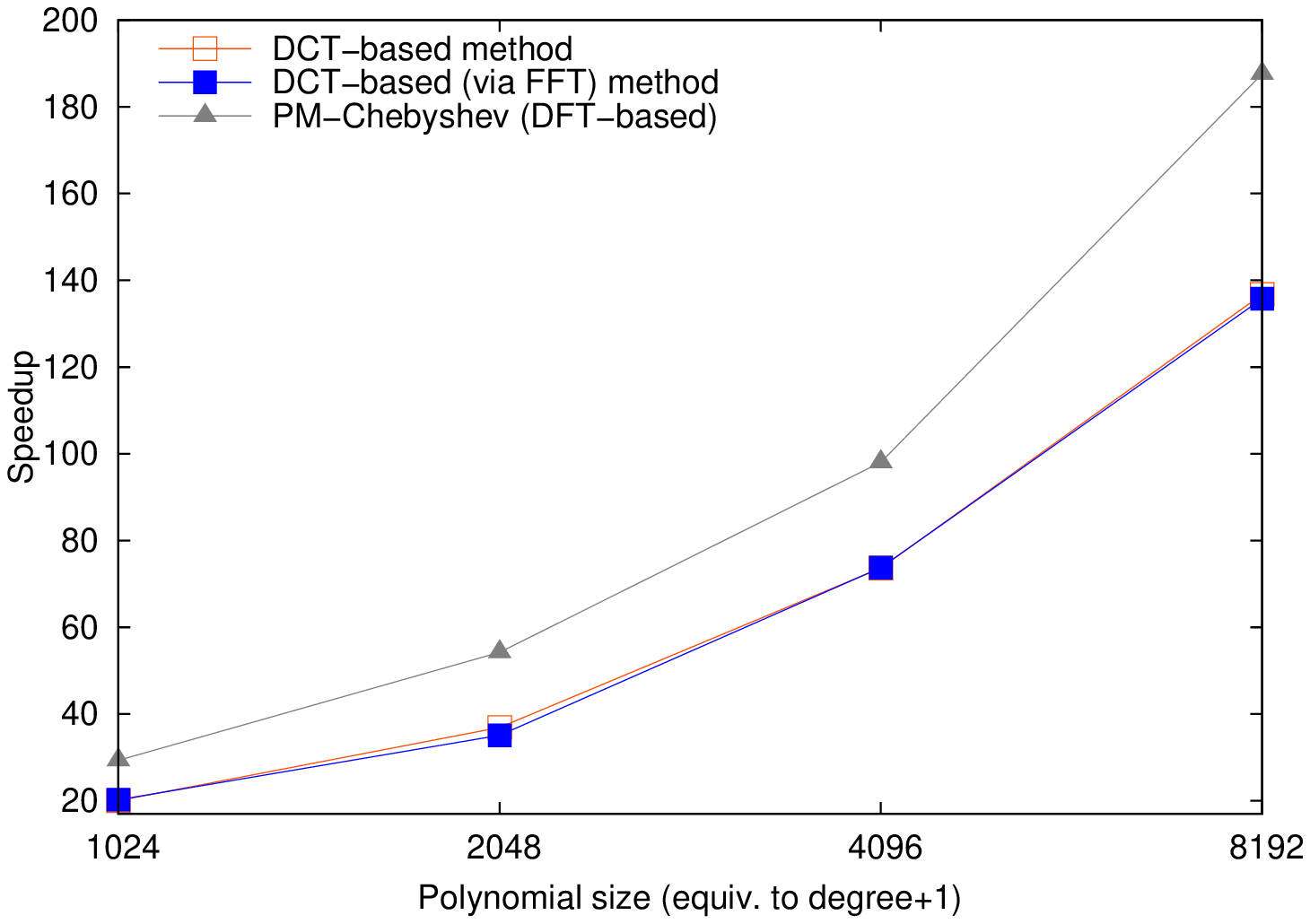}
\end{figure}

We report in Figure \ref{fig:perf_intel}  the relative performances to the Direct method implementation for polynomial sizes ranging from 2 to 8192. Both axis use logarithmic scale, and the ordinates axis represents the speedup against Direct method. All times used in this figure are given in Table \ref{table:perf}. One can also find in Figure \ref{fig:partialview} more detailed views of the Figure \ref{fig:perf_intel}.
As expected, one can see on these Figures that the Direct method reveals the most efficient for very small polynomials (i.e. polynomial degrees less than 16). This is explained by the low number of operations required by this method and its simplicity which makes possible several optimizations by the compiler (e.g. loop unrolling).  
When polynomial sizes are getting larger, the methods based on discrete transforms become the most efficients.
In particular, we can see that DCT-based method is catching up with its version based on FFT, which clearly illustrates that DCT-I implementation of FFTW is using a double length FFT, as explained in Section \ref{sec:dctdft}. 
Therefore, as expected, our Algorithm {\tt PM-Chebyshev} (DFT-based)  is the most efficient with polynomial sizes greater than 16. In particular, our {\tt PM-Chebyshev} (DFT-based) implementation is gaining on average $20\%$ to $30\%$ of  efficiency over the DCT-based implementations. This result almost satisfies the theoretical estimates since the complexity gain is asymptotically of $33\%$ (taking into account only the constants of high order terms in the complexity). 

\medskip
\begin{remark}
One could have been interested to see the practical behavior of the method of Lima {\it et al.} \cite{LPW2010}.
However, our feelings on the efficiency of such a method lead us to be pessimistic. Even if this method decreases the number of multiplications, 
it increases the overall number of operations. Moreover, this method needs an important amount of extra memory (i.e. $O(n^{\log 3})$) which definitively
increases data access and then should considerably penalize performances. Furthermore, the method is quite complex, especially for the indices management in  the separation procedure. Since no detailed algorithm is given in \cite{LPW2010} it is not easy to make an implementation and then offer a fair comparison.
Finally, from our benchmarks we observe that the performance of the Karatsuba multiplication does not compete with the Direct method for  small polynomials (e.g. size less than 16). Adding the storage of intermediate value within Karatsuba procedure plus the extra quadratic operations needed by the method of Lima {\it et al.} \cite{LPW2010} will probably make its implementation not competitive with other existing methods.
\end{remark}

\begin{table*}[H]
\begin{center}
\caption{Times of polynomial multiplication in Chebyshev basis (given in $\mu s$) on Intel Xeon 2GHz platform.}\label{table:perf}
\renewcommand{\arraystretch}{1.5}
\begin{tabular}{|c||r|r|r|r|r|}
\hline
\bf n & \bf Direct  & \bf DCT-based  & \bf DCT-based (FFT) & \bf PM-Cheby (Kara) & \bf PM-Cheby (DFT)\\
\hline 
   2 &      0.23 &    1.25 &    0.43 &     0.41 &    0.49\\
   4 &      0.43 &    1.32 &    0.57 &     0.62 &    0.58\\
   8 &      0.52 &    1.80 &    0.81 &     1.04 &    0.86\\
  16 &      1.28 &    3.11 &    1.56 &     1.91 &    1.28\\
  32 &      4.33 &    4.88 &    2.84 &     5.11 &    2.58\\
  64 &     15.73 &    8.82 &    9.55 &    16.83 &    5.41\\
 128 &     56.83 &   18.08 &   17.07 &    70.54 &   13.37\\
 256 &    211.34 &   35.97 &   31.42 &   195.85 &   24.41\\
 512 &    814.71 &   82.47 &   68.67 &   554.36 &   49.53\\
1024 &   3219.13 &  160.81 &  159.21 &  1618.37 &  109.74\\
2048 &  12800.30 &  346.95 &  364.82 &  4749.63 &  236.01\\
4096 &  54599.10 &  741.38 &  740.31 & 14268.10 &  556.83\\
8192 & 220627.00 & 1613.43 & 1625.43 & 40337.20 & 1176.00\\
\hline
\end{tabular}
\bigskip

{\bf PM-Cheby} stands for {\tt PM-Chebyshev} algorithm.
\end{center}
 \end{table*}

\section{A Note on Problems Equivalence}\label{sec:equivalence}

Let us consider the problem of multiplying two polynomials given by their coefficients in a given basis of the $\R$-vector space of $\R[x]$.
We denote this problem in monomial basis as $M_{mon}$ and the one in Chebyshev basis as  $M_{che}$.
Under this consideration, one can demonstrate the following theorem:\\

\begin{thm}
Problems $M_{mon}$ and $M_{che}$ are equivalent under a linear time transform,  $M_{mon} \equiv_L M_{che}$, and the constant of both transforms is equal to two.
\end{thm}
\medskip
\begin{proof}
As we have shown in Section \ref{sec:reduction} the problem of multiplying polynomials in Chebyshev basis linearly reduces to the multiplication in monomial basis, and the constant in the reduction is two. Thus we have already demonstrate $M_{che} \leq_L M_{mon}$.

We can  show that $M_{mon} \leq_L M_{che}$ by using (\ref{eq:directmethod2}).
Indeed, we can see from (\ref{eq:directmethod2}) that the $d+1$ leading coefficients of the product in Chebyshev basis exactly match with the ones in monomial basis on the same input coefficients. It is easy to show that the remaining $d$ coefficients can be read from the product in Chebyshev basis of the reversed inputs. 
Let us denote $\times_{c}$ the multiplication in Chebyshev basis and $\times$ the one in monomial basis.
Consider the two polynomials $\bar{a},\bar{b}\in \R[x]$  given in monomial basis as
\[
\displaystyle \bar{a}(x)=\sum_{k=0}^d a_kx^k \mbox{ and } \bar{b}(x)=\sum_{k=0}^d b_kx^k.
\]
Consider the  polynomials $a(x),b(x),\alpha(x)$ and $\beta(x)$ sharing the same coefficients as $\bar{a}(x)$ and $\bar{b}(x)$ but expressed in Chebyshev basis:
\[
\displaystyle a(x)=\sum_{k=0}^d a_kT_k(x) \mbox{\, ,\, } b(x)=\sum_{k=0}^d b_kT_k(x),
\]
\[
\displaystyle \alpha(x)=\sum_{k=0}^d a_{d-k}T_k(x) \mbox{\, ,\, } \beta(x)=\sum_{k=0}^d b_{d-k}T_k(x).
\]
The coefficients $c_k$ of the polynomial $\bar{c}(x)=\bar{a}(x)\times \bar{b}(x)$  expressed in monomial basis
can be read from the coefficients of the polynomials
\[
f(x) =a(x)\times_c b(x) \mbox{ and } g(x) =\alpha(x) \times_c \beta(x)
\]
using the  relation
\[
c_k=
\begin{cases}
g_{d-1-k} & \text{for } k=0\hdots d-1,\\
f_k       & \text{for } k=d\hdots 2d.
\end{cases}
\]
This clearly demonstrates that $M_{mon} \leq_L M_{che}$ and thus complete the proof.
\end{proof}
\section{Conclusion} 

We described yet another method to reduce the multiplication of polynomials given in Chebyshev basis to the multiplication in the monomial basis.
Our method decreases the constant of the problem reduction and therefore offer a better complexity than the ones using basis conversions.
Moreover, since our method does not rely on basis conversions, it might offer more numerical stability as it could be when converting 
coefficients to other basis. As we already mentioned, the problem of numerical stability is of great interest and should be treated as a dedicated article.

Our {\tt PM-Chebyshev} algorithm offers an efficient alternative to any existing quasi-linear algorithms.
In particular, it allows to use  Fast Fourier Transform of half length of the one needed by the specialized DCT-based method, which is an alternative when DCT codes are not available or sufficiently efficient. In such a case, our method achieves the best performances among all the available method for large degree polynomials.
 
Finally,  our attention in this work has been focused only on polynomials in $\R[x]$ using Chebyshev basis but our approach is still valid for other basis related to the Chebyshev one and other domains. For instance, our method will work for polynomials over finite fields using a basis of Dickson polynomials since they are a generalization of the Chebyshev polynomials (see \cite{FF1997}). More generally, our method will work for any basis and any domains satisfying (\ref{eq:chebyprod}), and any variant relaxing the constant factor.

Although our reduction scheme using Karatsuba's method is not as efficient as one could have expected for polynomials of medium size, 
further work to optimize its implementation should be investigated.
 This is of particular interest since such medium size polynomials are used in validated computing to approximate functions using a Chebyshev Interpolation model \cite{BJ2010}.
This has also a practical interest since Chebyshev or Dickson polynomials can be used in cryptography applications \cite{HasanNegre2010,LimaPanarioSouza2010}, which often need medium size polynomials for defining extension field (e.g. $\mathbb{F}_{2^{160}}$).

One possible way to optimize our Karatsuba based method is to reduce the number of additions by modifying our reduction as we did for the FFT-based approach in section \ref{sec:dctdft}. Indeed, since we use almost the same operands within the two Karatsuba's multiplications one can save almost half of the additions involved in adding the higher and the lower parts of each operand. It seems also feasible to apply this saving along the recursive calls of Karatsuba method.
Another interesting way to improve the efficiency would be to provide an implementation of Karatsuba algorithm minimizing the extra memory as the one proposed in \cite{Harvey:2010}.

\section*{Acknowledgment} \addcontentsline{toc}{section}{Acknowledgment}
The author would like to thank Claude-Pierre Jeannerod, Laurent Imbert and Arnaud Tisserand for their helpful comments and suggestions during the preparation of this article. The author is also grateful to the anonymous referees, especially for suggesting the numerically stable optimization of the Algorithm {\tt PM-Chebyshev} (DFT-based).

\end{document}